\documentclass[12pt,a4paper]{article}
\usepackage[latin1]{inputenc}
\usepackage{graphicx}
\usepackage{amsmath}
\usepackage{cite}
\usepackage{multirow}
\title{Magic Pore Dynamics in Weakly Interacting Clusters}
\author{$^{1}$M.A.Ratner, $^{1,2}$V. V. Yanovsky}
\begin{document}

\maketitle

$^{1}$\textit{Institute for Single Crystals NASU, 60 Nauki Ave, 61001 Kharkiv, Ukraine}

$^{2}$\textit{Kharkiv National University, Svobody Sq. 4, 61000 Kharkiv, Ukraine}

\abstract{Process of relaxation of Lennard-Jones cluster with an intrinsic pore was investigated by molecular dynamics method for different phase states of an initial cluster. Strong dependence of pore relaxation character on the initial cluster phase state was demonstrated. It was shown that, for the initially solid cluster, the system can reach metastable state where pore radius is fixed at one of the discrete set of values. These values do not depend on initial cluster temperature and size. Thus, it was demonstrated that, in a wide range of cluster sizes, inside solid clusters "magic" pores can stabilize forming a metastable state.}


\section{Introduction}

It is well known that properties of materials are to a large degree determined by their defect structure. Various defects of materials can be classified by their dimensionality. Thus, pores and new phase inclusions relate to 3-dimensional defects, while grain boundaries relate to 2-dimensional and dislocations to 1-dimensional ones. It is naturally to classify such elementary objects as vacancies and atoms in interstitials as 0-dimensional defects. If estimated according to the effect on many material properties, 3-dimensional defects, such as pores and new phase inclusions, play the key part. In the bulk materials, where the role of boundaries is insignificant, theory of behavior and evolution of such objects is well developed (see e.g. \cite{Slezov,SchmelzerSlezov,SchmelzerUlbricht,Kukushkin}). In these works the main attention was paid to collective behavior of an ensemble of pores. However, in connection with intensive investigation of nanomaterials, where the influence of boundaries is enormous, it came out, that, in such materials, the behavior of pores and new phase inclusions is not sufficiently investigated. It must be noted that the problem of pore evolution inside a nanoparticle is significantly more complicated than in macroscopic materials. Firstly, it is close to the problem of the interaction of two pores \cite{DubinkoRus}. Here the role of the second participant is played by the pore boundary. Secondly, the search for anomalous pore behavior in such particle requires comparison with classical pore behavior on the account of diffusion fluxes. Evolution of pore in nanoparticle in diffusion approximation was considered in details in \cite{OurPaper}. Here nonlinear differential equations are obtained for the evolution of pore radius and position and all possible asymptotic regimes of pore behavior are considered. It is shown that vacancy pore in the process of evolution is diminishing and moving towards the center of nanoparticle. In the present work we concentrate on the distinctions of pore evolution in a nanoparticle that arise at more detailed consideration that exceeds the frame of diffusive approximation. Small clusters with pair interactions were investigated in details with the use of numerical modeling \cite{Berry, Smirnov2007, Smirnov2000, Smirnov1994, Wang}. However, cluster sizes in these simulations are too small to include an intrinsic pore. In the work \cite{Berry}, it was demonstrated numerically for clusters of small sizes, that high configuration excitations of a cluster correspond to formation of cavities (agglomerates of several vacancies) inside it. As transition of a cluster occurs from a liquid state to solid one, the cavities go out of the cluster. Such process, related to diffusion of cavities towards cluster surface, is of activation (barrier) character.

In the present work, the relaxation of a cluster with an intrinsic pore was investigated by the molecular dynamics method that allows to take into account the discrete intrinsic building of the cluster and consider clusters in different phase states. The distinction of the present work from the previous ones is that large enough clusters that can include an intrinsic pore of several lattice periods are modeled by MD method. As it is demonstrated below, the phase state of the initial clusters influence greatly the character of the relaxation of small cluster with an intrinsic pore. Thus, it is shown that inside solid clusters "magic" pores can stabilize forming a metastable state. This phenomenon is similar to "magic" clusters that are more stable than other ones, with the difference that role of atoms here is played by "closely packed" vacancies. Evidently, such pore structure minimizes the number of bonds that are broken as a result of pore formation.

\section{Model and method}\label{sec:modmeth}

\subsection{Modeling parameters}\label{ss:modpar}

In the present work, modeling of pore relaxation in a nanoparticle was carried out by molecular dynamic method that allows to consider different phase states of the initial cluster and their influence on pore relaxation process. The object of the study were Ar clusters with number of atoms up to 10648 ($22^3$). Rare gas clusters are one of the simplest systems, in particular three-particle interactions can be neglected in this case \cite{Berry}. Up to the size of several thousand atoms, rare-gas clusters form icosahedral structure, that allows to minimize cluster surface, for larger clusters transfer to FCC structure occurs.

Initially, cubic lattice of Ar atoms (atomic mass \textit{m}=39.9 a.u.) was constructed with the lattice period $a$ = 4.816 \r{A} (corresponding to solid Ar). Atoms are interacting via Lennard-Jones potential with the following parameters, taken for Argon atoms: $\sigma$=3.405 \r{A}, $D$=0.01032 eV, $R_{cut}=2.5r_{min}$. Here \textit{D} is the depth of the potential well, $\sigma$ is the finite distance at which the inter-particle potential is zero, $R_{cut}$ is cut-off distance. The inter-atomic force and its derivative was smoothed at the cut-off distance. Cluster temperature \textit{T} is defined as mean kinetic energy per one atom.  Below, the temperature is measured in the units of potential depth \textit{D} while linear sizes are measured in the units of equilibrium inter-atomic distance for a pair Lennard-Jones potential, $r_{min}=2^{1/6}\sigma$.

The equations of motion are solved numerically via velocity Verlet algorithm \cite{Verlet, Andersen} with a time step $dt = 10^{-6} \div 10^{-5}$ps, dependently on cluster temperature.

Pore relaxation is described in a following way. The whole cluster is overlaid with a three-dimensional cubic lattice with  a period $a_{grid}=0.5r_{min}$. A pore is defined as any connected domain of cells, that are free from cluster atoms, with a maximal diameter no less then $2.5r_{min}$, and surrounded on all sides by occupied cells. If such domain has maximal diameter less than $2.5r_{min}$ but greater then $2r_{min}$, it is defined as vacancy. Such criterion was obtained in empirical way in order to discern pore with single vacancies and avoid the influence of surface convolution. Cluster volume is found as summarized volume of occupied cells and free cells, belonging to pores or vacancies.

\subsection{Preparation of initial cluster conformations}\label{ss:prep}

As it will be shown below, relaxation of a cluster with an intrinsic pore depends strongly on the initial cluster phase state. Thus, equilibrium clusters in different phase states should be prepared in the investigated range of temperatures.

\subsubsection{Phase state definition}\label{sss:phase}

In order to define cluster phase state, the fluctuation $\delta$ was used of square root of mean square distance between neighbor atoms, $a$. According to the conventional criterion \cite{Smirnov2007}, for solid state
\begin{equation}
\delta < 0.15a
\label{eq1}
\end{equation}
Phase states of small (up to several thousands atoms) clusters with pair interactions have some specific features (see,e.g.\cite{Smirnov2007},\cite{Berry}). Origin of these peculiarities lies in the internal structure of such clusters that consist of consequent icosahedral shells. This way number of broken bonds on cluster surface is minimized. If all icosahedral shells are completely filled, the number of cluster atoms is called "magic". Such "magic" clusters are more stable than other ones; various cluster properties, such as melting temperature and heat capacity as functions of cluster size are unusually high for magic clusters.

At low enough temperatures clusters are in solid state where criterion (\ref{eq1}) is met throughout the entire cluster from the core to the surface.

As the temperature of solid cluster rises, criterion one becomes violated for its outer shell (i.e.it is melted), since number of bonds is smaller for outer atoms. As cluster temperature continues rise, the next shell before outer is melted.  Thus, cluster melting occurs from its outer shells towards its core.

Thus, second type of the cluster phase state can be defined where criterion (\ref{eq1}) is met for the entire cluster except for one or several outer shells. As it was mentioned in \cite{Smirnov2007}, the second type of phase state presents a set of close phase states, each corresponding to different number melted outer shells.

As cluster temperature continues rise, all cluster shells become melted. The liquid state of a cluster is the least ordered state where criterion (\ref{eq1}) is violated throughout the entire cluster.

\subsubsection{Procedure of obtaining clusters in different phase states}\label{sss:obtain}

Initial state of a cluster, constructed via procedure described in Sec. \ref{ss:solid} is far from equilibrium and, if allowed to relax freely, finishes in liquid state with high enough temperature. The time between phase transitions of cluster, in the researched  range of cluster sizes and temperatures, is too large to be observed during numerical count. In the present investigation, clusters in solid state are obtained in the following way. After assigning initial atom coordinates, cluster is quickly cooled down to temperature 20 K\textit{ }(that is below melting temperature). This way, cluster in a liquid (overcooled) state is always obtained, since the system has not enough time to relax to the solid state, even if it is equilibrium. Then, cluster is continued to be cooled down by quasi-adiabatic method to temperature $T_{min}$ (decrease temperature by 2\% every 0.1 ps), that is close to zero. This way, transfer of the cluster into solid state is forced. After that, cluster is heated by the same quasi-adiabatic procedure until it reaches required temperature $T_0$. This way we obtain a cluster in solid state that is characterized by the lowest specific potential energy (potential energy per atom), $U_{s}$(see Fig. \ref{fg1}). Besides obeying criterion (\ref{eq1}), this cluster is characterized by close packing of atoms that is typical for small clusters with pair interactions of atoms \cite{Berry, Smirnov2007, Smirnov2000}. If $T_{min}$ is increased, the less ordered state with solid core and liquid shell is obtained, that is characterized by higher $U_{s}$.

\subsection{Potential energy of obtained clusters in different phase states}\label{ss:energy}

\begin{figure}[h]
\centering
\includegraphics[width=12 cm]{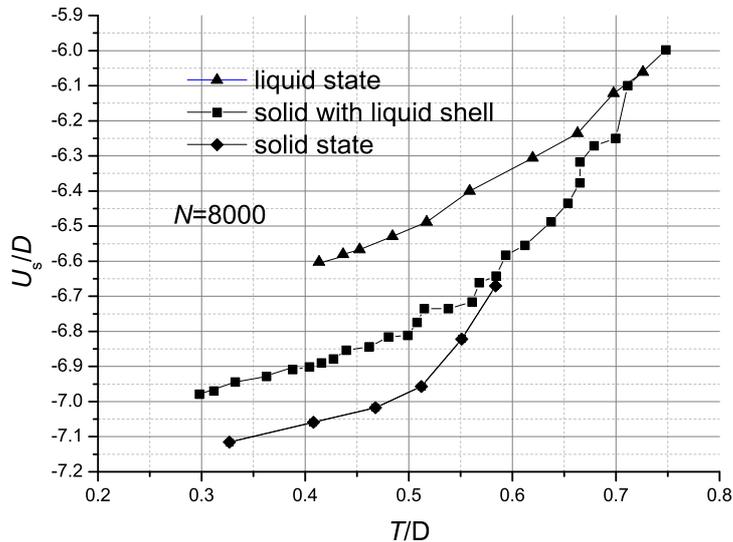}
\caption{The examples of the dependence of specific potential energy on cluster temperature for various stable (metastable states) for $N$=8000. Rhombuses indicate solid state, squares -- solid core with liquid shell while triangles relate to liquid state. For each point of the plot, the results are averaged over\textit{Q}=25 random cluster realizations}
\label{fg1}
\end{figure}

In Fig. \ref{fg1}, the example (for $N$=8000) is given of the dependences of specific potential energy $U_s$ on cluster temperature $T$. These dependences are shown for cluster in different phase states: solid, liquid and the one with solid core and liquid outer shells. It can be seen from the figure, that, at the same temperature, the cluster can have different potential energies that correspond to different phase states (the more ordered is cluster state, the lower is its potential energy). Evidently, at the given temperature, one of these phase states is stable, while the others are metastable one. However, the simulation time is too short for the observation of transitions between stable and metastable states and obtaining corresponding statistics (as it was mentioned in \cite{Berry, Smirnov2000}, hysteresis phenomenon is essential for small clusters and reveals itself both in computer simulations and in experimental studies). Thus, within given investigations, relaxation of clusters with an intrinsic pore was conducted separately for solid clusters and for clusters with solid core and liquid shell(s) without studying phase transitions between those states. Still, authors believe that the given work reveals some essential features of the relaxation of cluster with intrinsic pore.

The fact, that melting of small cluster is realized via consequent melting of its outer shells \cite{Smirnov2007},\cite{Berry}, can explain the existence of "footsteps" in Fig. \ref{fg1} for the dependence of $U_s$ on $T$ for solid cluster with liquid shell (shells). In the further consideration, by the melting temperature $T_{melt}$, we will mean the temperature at which all cluster shells are melted.

\subsection{Obtaining initial random conformations of clusters with an intrinsic pore}\label{ss:relax}

Equilibrated clusters at required temperatures in different phase states are obtained by the described above method. Then, for each cluster with given temperature, MD run is started. Every 5 ps one conformation is fixed that is included in the set of random conformations at the given temperature. This way, for a cluster with a given $N$, $T$, the number $Q=25$ of random realizations of initial cluster is created (with the mean velocity corresponding to the given temperature $T$). For each of obtained initial cluster realizations, a pore of the given initial radius $R_{pore}(0)$ is instantly cut off in the center of a cluster. After that, the numerical solving of motion equations is carried out by  Verlet algorithm (with parameters given in Sec. \ref{ss:modpar})  in order to investigated system relaxation.

\section{Results and discussion}\label{sec:results}

In the present work, Ar clusters with number of atoms  $N=14^3$, $16^3$, $18^3$, $20^3$, $22^3$ were investigated in the temperature range 0.36D-0.80D. Clusters in solid state as well as in solid state with liquid outer shell we considered. $Q=25$ of random realizations of equilibrated clusters for each investigated pair ($N$, $T$) were obtained by the described in the previous section method, over which the results were averaged.  For each of obtained initial cluster realizations, a pore of the given initial radius $R_{pore}(0)$ was instantly cut off in the cluster center. Then MD run was conducted for about 30 ps  in order to investigated system relaxation. For all clusters Investigated pore radii were  $R_{pore}(0)=3r_{min},4r_{min}$, for $N=22^3$, pore radius $5r_{min}$ was also investigated.

\begin{figure}[h]
\centering
\includegraphics[width=12 cm]{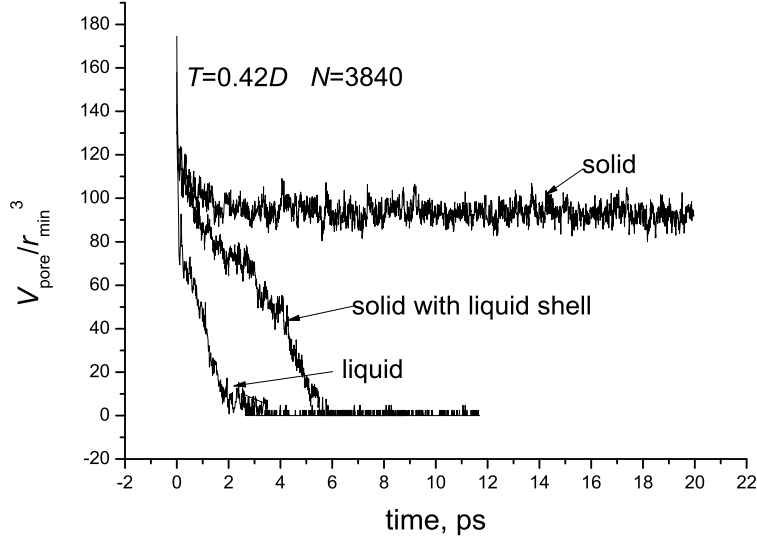}
\caption{The examples (random realizations) of relaxation of cluster in a solid state and in a state with solid core and liquid shell ($N=3840$, $T=0.44D$, $R_{pore}(0)=3r_{min}$). The typical dependence of pore relaxation character on initial cluster state is demonstrated.}
\label{fg2}
\end{figure}

Molecular modeling results demonstrate that pore healing process depends drastically on the initial cluster phase state. Thus, in Fig. \ref{fg2}, the example is given for the time dependences of pore volume for a cluster in liquid state, solid state with a liquid shell and  completely solid state (determined according to criterion (\ref{eq1})). The dependences are given for initial cluster size $N=3840$, initial pore radius $R_{pore}(0)=3r_{min}$, $T=0.42D$ (45 K). It can be seen from Fig. \ref{fg2}, that in all cases, quick diminishing of pore volume occurs at first due to strongly nonequilibrium conditions. Then, much slower processes follow. For the liquid state and the solid state with liquid core, pore volume diminishes monotonously, while for the completely solid state of the cluster, pore relaxation is of barrier character. This effect was already described in the work \cite{Ratner}, where it is supposed that a solid cluster with an intrinsic pore can present a metastable configuration, separated from a stable state by energy barrier. In this case, the preferable pore sizes exist, that can persist for a relatively long time.

\subsection{Pore relaxation in the case of solid cluster}\label{ss:solid}

Firstly, we should mention that cutting pore out of a cluster in the described above way gives rise to a nonequilibrium process that can change cluster phase state. For each cluster size, there exists critical temperature $T_c$, below which, cluster remains solid after cutting pore inside it. The values of $T_c$ are given in Table \ref{t:tab1} dependently on cluster size. Evidently, $T_c$ is close to the temperature of cluster phase transition between completely solid state and solid state with liquid outer shell. The exact determination of the temperature of such transition by means of MD is hindered by hysteresis phenomenon.

\begin{figure}[h]
\includegraphics[width=7 cm]{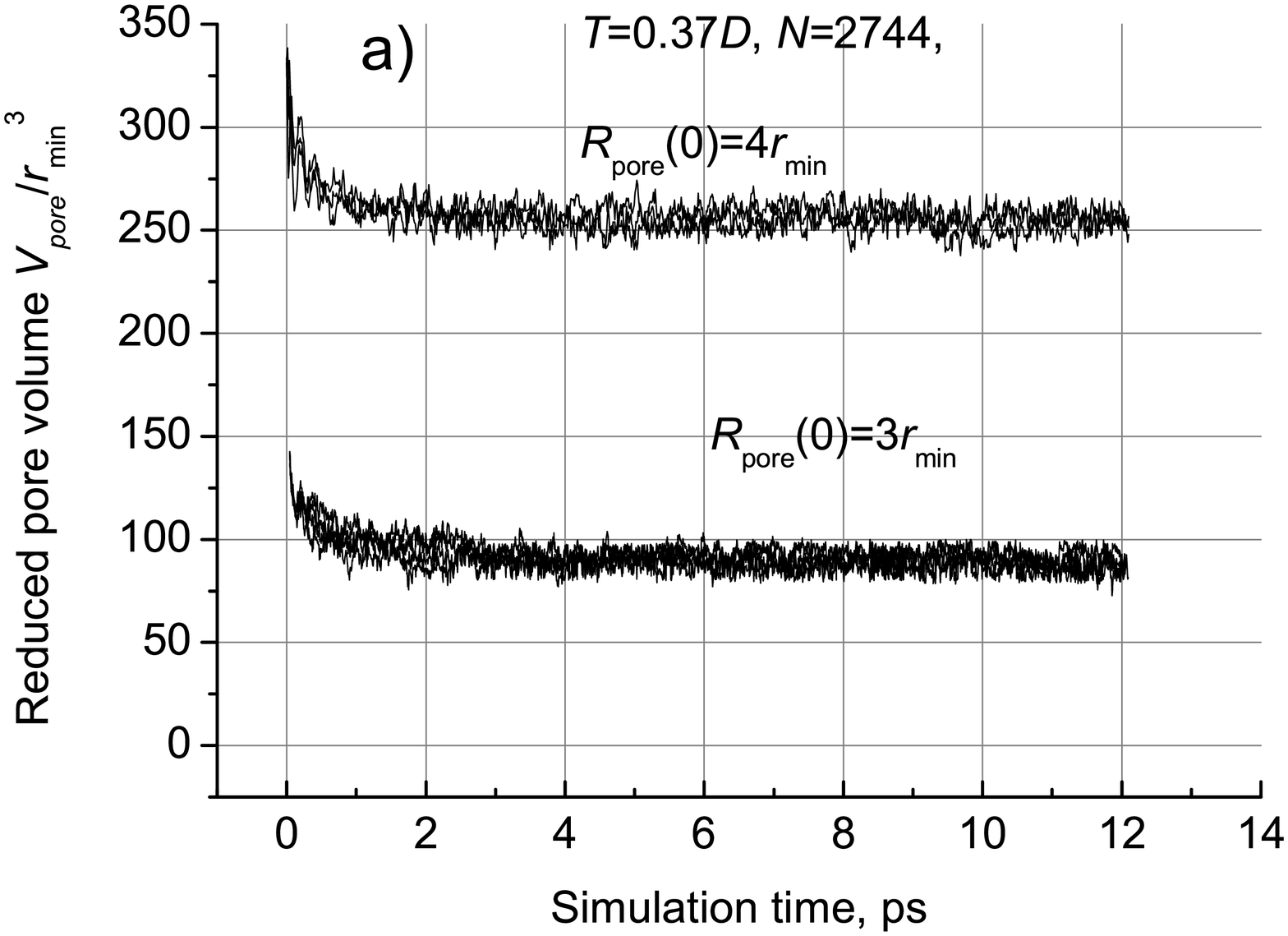}
\includegraphics[width=7 cm]{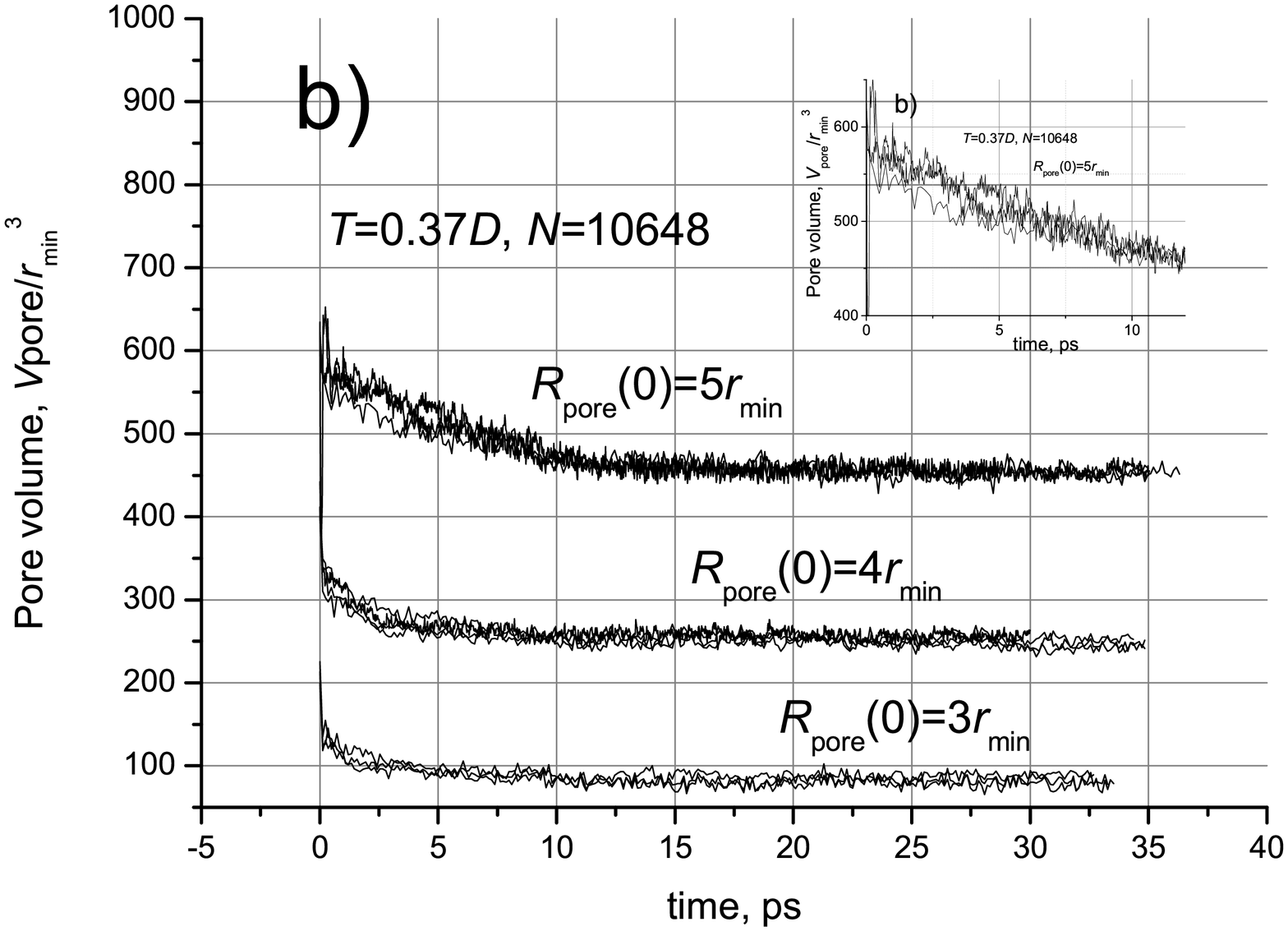}\\
\includegraphics[width=7 cm]{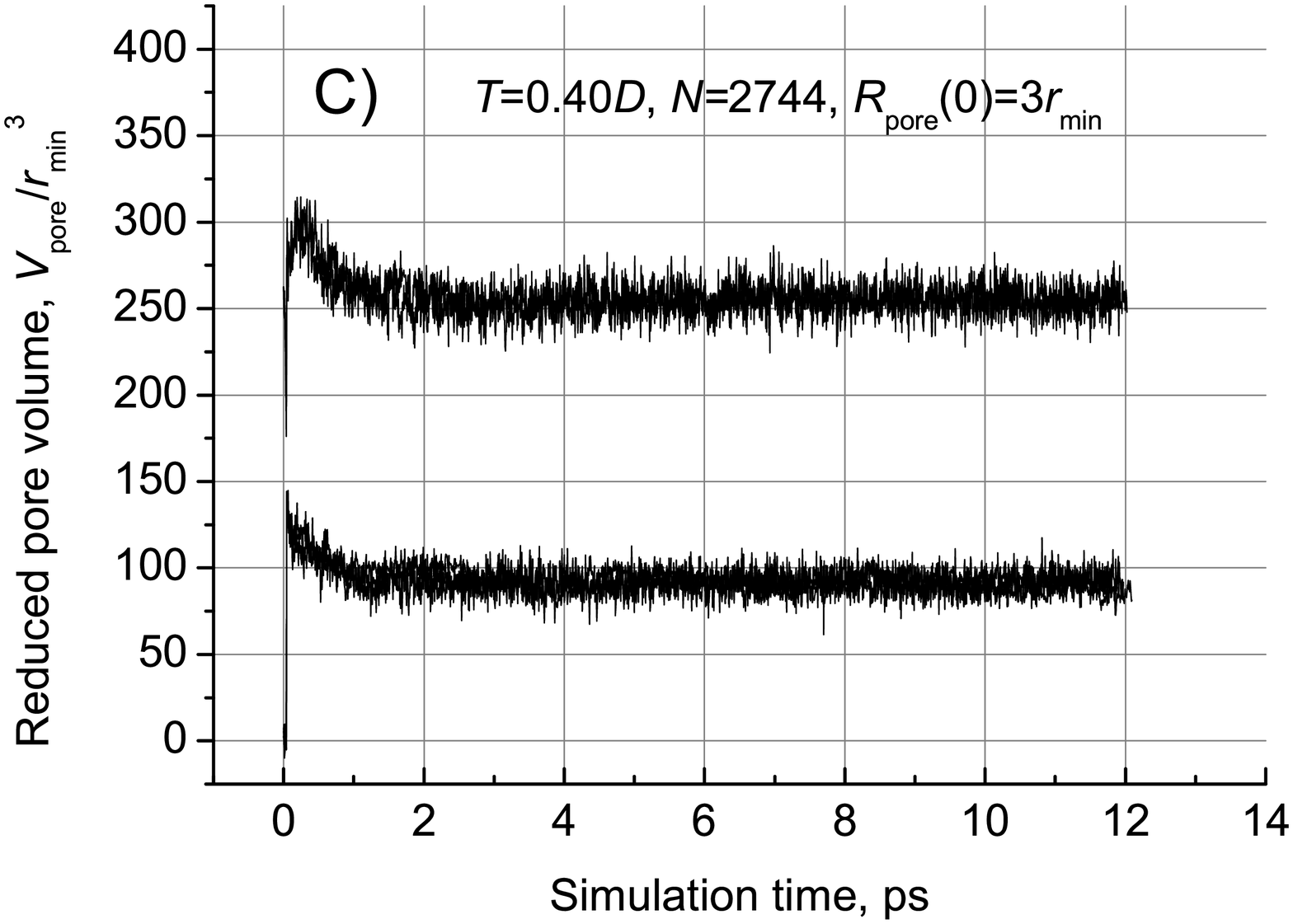}
\includegraphics[width=7 cm]{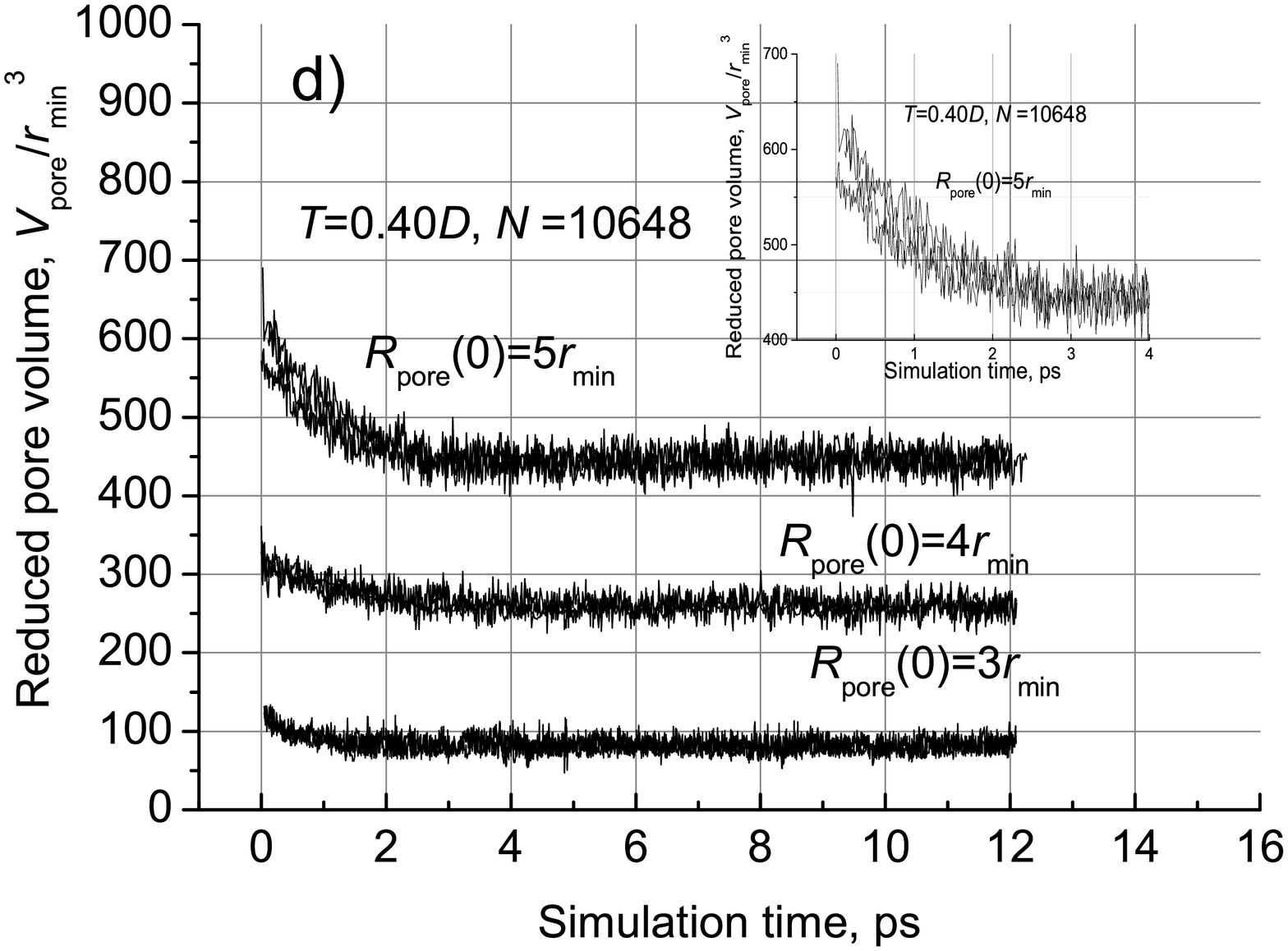}
\caption{The examples (random realizations) of relaxation of a cluster in a solid state at $T=0.37D$, $0.40D$, \textit{N}=2744, 10648, $R_{pore}(0)=3r_{min}$, $4r_{min}$, $5r_{min}$.}
\label{fg3}
\end{figure}

Examples of time dependence of pore volume in clusters of various sizes below $T_c$ are shown in Fig. \ref{fg3}a)-d).

In the present investigation, relaxation of the pores with initial radii $R_{pore}(0)=3r_{min}$ and $4r_{min}$ was investigated below $T_c$ for the clusters containing $N=14^3$, $16^3$, $18^3$, $20^3$, $22^3$ atoms. For the cases $N=22^3$, relaxation of pores with initial radius $R_{pore}(0)=5r_{min}$ was additionally investigated. In all these cases, pore volume, irrespectively of initial cluster size and temperature, after initial quick decrease, fixes itself at certain value $V_{stable}$ that depends only on the initial pore size. The examples of such dependences are shown in Fig.\ref{fg3} for initial cluster sizes $N=14^3$, $22^3$ and temperatures $T=0.37D$, $0.40D$. The values of $V_{stable}$ are shown in the Table \ref{t:tab1}. Each value of stabilized pore volume $V_{stable}$ can be associated with the number of "missing" closely packed atoms, or vacancies in the pore $N_{vac}=V_{stable}/r_{min}^32^{1/2}$. Values of $N_{vac}$ for different initial pore sizes are  close to the numbers of atoms in icosahedral atomic clusters with closed shells (so-called "magic" clusters). Thus, for $R_{pore}(0)=3r_{min}$, $N_{vac}\sim 136$ that is close to number of atoms in  icosahedral cluster with 3 closed shells (147 atoms), for $R_{pore}(0)=4r_{min}$, $N_{vac}\sim 354$ that is close to the number of atoms of icosahedral cluster with 4 closed shells that equals to 309.  As it was indicated above, atomic clusters of such "magic" sizes demonstrate unusual physical properties, in particular, high stability, see e.g. \cite{Berry, Smirnov2007, Smirnov2000, Smirnov1994}). The above said allows us to speak about "magic" pores (or vacancy clusters) that demonstrate unusually stable properties in the way similar to "magic" clusters. Like the case of atomic cluster, the unusual vacancy cluster stability is due to an optimal surface (interface) shape that minimizes number of broken bonds.

Let us underline, that value of stabilized pore volume $V_{stable}$ practically does not depend on cluster temperature and initial cluster size, but only on initial pore size (see Table \ref{t:tab1}).

The energy barrier between metastable state with a pore and stable state without one is, probably, overcome due to vibration of the system as a whole (collective motions of atoms, so called breathing mode), that was observed in the present modelling as well as in a number of experimental and theoretical works on nanoparticles \cite{Mankad, Prafulla}.

\subsection{Pore relaxation in a cluster with solid core and liquid shell}\label{ss:liquid}
\begin{figure}[h]
\centering
\includegraphics[width=12 cm]{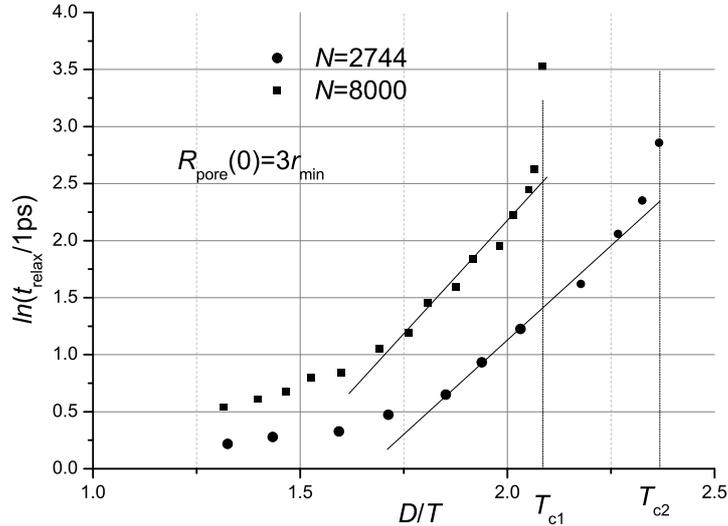}
\caption{Typical dependences of natural logarithm of pore relaxation time $t_{relax}$ on reverse cluster temperature $D/T$ are presented on the examples of $N=8000$ and $2744$. Straight lines indicate linear approximation with coefficients $U_{act}$ presented in Table \ref{t:tab1} (error within 5\%). Vertical dashed lines indicate points where temperatures become critical ($T_{c1}$ for $N=2744$ and $T_{c2}$) for for $N=8000$. Each point in the plot is averaged over \textit{Q}=25 random cluster realizations }
\label{fg4}
\end{figure}

As it was mentioned above, cluster melting begins from is outer shell, then next shells are consequently melted.
As shows present simulations, at $T>T_c$ cluster interface with a pore is also melting. Thus, in this case, we can not expect to obtain stable pore as in case of solid cluster. Indeed,  at $T>T_c$ pore relaxation process accelerates sharply (as it is illustrated in Fig. \ref{fg2} ). As examples, in Fig. \ref{fg4}, the dependences are shown of pore relaxation time $t_{relax}$ on reverse temperature $D/T$ for the cluster with solid core and liquid shells with $N$=1728 and 8000 and initial pore radius $R_{pore}(0)=3r_{min}$.

As initial cluster temperatures rise above the critical one ($T>T_c$), these dependences become linear with proportionality coefficient $U_{act}$ given in Table \ref{t:tab1}. This corresponds to the vacancy diffusion from pore surface towards the cluster surface, obeying Arrhenius low. Then Einstein relation gives
\begin{equation}
t_{relax} \sim exp(-\frac{U_{act}}{T})
\label{eq3}
\end{equation}
The same result is obtained in \cite{OurPaper} for the continuous model of cluster with a pore for large enough pore radii.
Here diffusion activation energy  $U_{act}$ corresponds to several $D$ (see Table \ref{t:tab1}).  As initial cluster temperature continues to rise,  pore relaxation time dependence on initial cluster temperature becomes essentially less steep. This is evidently connected with consequent melting of inner cluster shells. Let us note that the temperature $T_{melt}(N)$ of complete cluster melting (transferring to a liquid state), obtained within present modelling, is probably higher than the real phase transition temperature for the given cluster size due to hysteresis phenomena \cite{Berry, Smirnov2007}.

\begin{table}[h]
\caption{Values of system relaxation parameters for various initial cluster sizes $N$:stabilized dimensionless pore volumes $V_{stable}/r_{min}^{3}$ (for solid cluster) at various initial cluster temperatures $T$,  cluster melting temperatures $T_{melt}/D$, characteristic vacancy diffusion activation energy for solid cluster with melted outer shell $U_{act}/D$ . For comparison: melting temperature of macroscopic solid Argon $T_{melt}=83.4$ K$ = 0.78D$}
\centering
\begin{tabular}{|l|l|c|c|c|c|} \hline
\multicolumn{2}{|l|}{$N$} & $14^3$ & $18^3$ & $20^3$ & $22^3$ \\ \hline
\multicolumn{2}{|l|}{$T_c/D$} & 0.42 & 0.46 & 0.48 & 0.49 \\ \hline
\multirow{2}{3.2cm}{$V_{stable}/r_{min}^{3}$, $R_{pore}(0)=5R_{min}$} & $T=0.37D$ & & & & 445 \\ \cline{2-6}
\ & $T=0.40D$ & & & & 447 \\ \hline
\multirow{2}{3.2cm}{$V_{stable}/r_{min}^{3}$, $R_{pore}(0)=4R_{min}$} & $T=0.37D$ & 251 & 256 & 254 & 255 \\ \cline{2-6}
\ & $T=0.40D$ & 253 & 254 & 258 & 257 \\ \hline
\multirow{2}{3.2cm}{$V_{stable}/r_{min}^{3}$, $R_{pore}(0)=3R_{min}$} & $T=0.37D$ & 88 & 94 & 92 & 93 \\ \cline{2-6}
\ & $T=0.40D$ & 93 & 95 & 94 & 95 \\ \hline
\multicolumn{2}{|l|}{$T_{melt}/D$} & 0.61 & 0.64 & 0.67 & 0.71 \\ \hline
\multicolumn{2}{|l|}{$U_{act}/D$} & 3.96 & 3.92 & 4.17 & 4.35 \\ \hline
\end{tabular}
\label{t:tab1}
\end{table}

\section{Conclusions}

The relaxation of a cluster with an intrinsic pore was investigated by the molecular dynamics method that allows to take into account the discrete intrinsic structure of the cluster.

It was demonstrated that character of the relaxation of small cluster with an intrinsic pore is determined by the initial cluster phase state. Thus, it was shown that, in a wide range of cluster sizes, inside solid clusters "magic" pores can exist for a relatively long time. The size of "magic" pores does not depend on initial cluster temperature and size, but only on initial pore size. Such system (cluster with an intrinsic "magic" pore) presents a metastable state. This phenomenon is similar to "magic" clusters that are demonstrate unusual properties, with the difference that role of atoms here is played by "closely packed" vacancies. Evidently, such pore structure minimizes the number of bonds that are broken as a result of pore formation.

For a cluster with solid core and liquid shell (or shells), pore healing process obeys Arrhenius diffusion law in a wide range of temperatures and corresponds to independent diffusion of vacancies from pore to cluster surface.

\end{document}